\documentclass{article}
\usepackage[preprint]{spconfa4}
\usepackage{graphicx}
\usepackage{booktabs}
\usepackage{amsfonts}
\usepackage{subfigure}
\usepackage{multirow}
\usepackage{bm}
\usepackage{hyperref}
\usepackage{amsmath}

\title{TRANSLATE REVERBERATED SPEECH TO ANECHOIC ONES: \\
SPEECH DEREVERBERATION WITH BERT}
\name{Yang Jiao}
\address{University of Maryland, College Park, Maryland, USA \\
yjiao1@umiacs.umd.edu}
  
\begin{document}
\pagestyle{plain}
%
\maketitle
\begin{abstract}
	Single channel speech dereverberation is considered in this work. Inspired by the recent success of Bidirectional Encoder Representations from Transformers (BERT) model in the domain of Natural Language Processing (NLP), we investigate its applicability as backbone sequence model to enhance reverberated speech signal. We present a variation of the basic BERT model: a pre-sequence network, which extracts local spectral-temporal information and/or provides order information, before the backbone sequence model. In addition, we use pre-trained neural vocoder for implicit phase reconstruction. To evaluate our method, we used the data from the 3rd CHiME challenge, and compare our results with other methods. Experiments show that the proposed method outperforms traditional method WPE, and achieve comparable performance with state-of-the-art BLSTM-based sequence models.
\end{abstract}
\section{Introduction}
\label{sec:intro}

In an enclosed space, reverberation is created when the sound wave propagates through multiple paths and superposes at the receiver \cite{naylor2010speech}. The reverberated speech is a convolution of the anechoic speech and impulse response of the enclosure The convolutional nature results in an expanded and smeared spectrogram. In contrast, speech denoising as a similar task, features additive background noise and direct superposition of spectrogram. Reverberation degrades the intelligibility of speech and poses a challenge for indoor applications of hearing aid devices and automatic speech recognition. There are a plethora of single channel and multichannel based statistical methods. A recent advancement is the application of a variety of neural network based methods. 

BERT model is first introduced in NLP for self-supervised language representation learning \cite{devlin2018bert}. Recently, BERT achieves success in speech representation learning as well \cite{chi2020audio,liu2020mockingjay,yang2020understanding}. BERT encodes contextual information via self-attention mechanism. Compared with Bi-directional Long Short-Term Memory (BLSTM), which consists of 2 separate unidirectional modules, the self-attention layer of BERT enables every token to attend to any other tokens in the sequence. In addition, computation of BERT is faster than BLSTM since it can be parallelized. Much deeper models can be trained for BERT than BLSTM. Therefore, it is natural in this work to compare BERT with BLSTM as the backbone sequence model for the task of speech dereverberation.

The adaptation of BERT model from NLP to speech is challenged by the noisy and continuous property of speech spectrogram. There are clear boundaries between discrete word tokens in NLP, whereas speech frames are locally smooth, change continuously across time and noisy in the sense that frames in a sequence are similar to each other \cite{chorowski2015attention}. In addition, developing positional encoding suitable for speech signal remains an open problem \cite{sperber2018self}. Therefore, in this work, we propose to apply a pre-sequence network to ameliorate those difficulties. We experimented with 5 variations: a default setup which applies a linear projection as in \cite{liu2020mockingjay}; 2 convolutional neural networks (CNN2d \& CNN 1d) which extract local temporal-spectral features; a (B)LSTM network in place of positional encoding to provide order information as in \cite{sperber2018self}; a CNN-LSTM (CL) net which enhances both local features and positional information. We compare BERT and BLSTM backbone models with 3 pre-sequence networks, CNN2d, CL and CNN1d. Hyperparameters are chosen to keep the model size similar. Our experiments show that BERT based methods achieve comparable performance with faster inference speed. 

Recent advancement in speech dereverberation and denoising features a plethora of self-attention based approaches. In \cite{liu2020mockingjay, chi2020audio}, BERT has been used for self-supervised speech representation which achieves success in a number of downstream tasks. In \cite{koizumi2020speech, hao2019attention, ge2019environment}, attention mechanism is used in combination with CNN and\slash or LSTM. However, in those works, modeling whole utterance as a sequence and learning hidden representations of speech robust to reverberation is not a major goal; attention mechanism is applied separately from a complete transformer layer. In \cite{kim2020t}, a transformer with Gaussian-weighted self-attention is proposed: the attention score between farther apart frames are attenuated. However, we don't apply any modification to the attention module: let pre-sequence network capture local features and let BERT capture high-level semantics. 

In addition to spectrogram reconstruction, phase reconstruction has been studied extensively in speech dereverberation and denoising. The approaches can be roughly divided into 3 categories: 1) reconstruct magnitude and phase of STFT separately, where our work falls into. 2) reconstruct STFT in complex domain \cite{williamson2015complex}. 3) reconstruct target speech in time domain \cite{pascual2017segan}. Most relevant to our work, \cite{ernst2018speech} estimates target spectrogram and reconstructs by using noisy phase and taking inverse ISTFT; \cite{kim2020t} reconstructs using Griffin-Lim algorithm, which takes the estimated magnitude and iteratively computes ISTFT that minimizes MSE with respect to input magnitude. In contrast, we use a pre-trained neural vocoder, Parallel WaveGAN \cite{yamamoto2020parallel}, which is a waveform generation model conditioned on input Mel-scale spectrogram. In our experiments, Parallel WaveGAN is about 10 times faster than Griffin-Lim while rendering comparable performance. 

\section{Proposed Method}
\label{sec:pagestyle}

\subsection{Problem Formulation}
\label{subsec: problem}

\setlength{\thickmuskip}{2mu} 
\setlength{\medmuskip}{1mu} 
\setlength{\thinmuskip}{0mu}

We consider a single channel speech dereverberation problem. The reverberated speech signal $y(t)$ is a convolution of room impulse response (RIR) $h(t)$ with clean speech signal $x(t)$ in the time domain. The dereverberation system produces $\hat{x}(t)$ as an estimate of $x(t)$. 

\begin{equation}
y(t) = h(t) \ast x(t)
\end{equation}

Let $Y(t,f)$ and $X(t,f)$ denote the short time Fourier transform (STFT) of $y(t)$ and $x(t)$. Let $H(t,f)$ denote the room transfer function (RTF). Then $Y(t,f)$ is obtained from a pointwise product of $X(t,f)$ and $H(t,f)$. 


\begin{equation}
Y(t,f) = H(t, f) \odot X(t,f)
\end{equation}

Let $M_x \in \mathbb{R}^{M \times T}$ and $M_y \in \mathbb{R}^{M \times T}$ denote the log Mel-scale spectrogram of $x(t)$ and $y(t)$. Let $C \in \mathbb{R}^{M \times F}$ denotes the Mel coefficients matrix. $M$ denotes the number of frequency bins in Mel scale, $F$ denotes the number of frequency bins at linear scale and $T$ denotes number of frames. $M_x$ is obtained by taking log of the matrix multiplication of $C$ and $|X|$. 

\begin{equation}
M_x=log(C \lvert X \rvert)
\end{equation}

In the proposed approach, a neural network takes $M_y$ as input and produces output $\hat{M_x}$ as an estimate of $M_x$. And then $\hat{x}(t)$ is produced by a neural vocoder conditioned on $\hat{M_x}$.

\subsection{System Overview}
   
\begin{figure}[tb]
\centering
\includegraphics[width=\columnwidth]{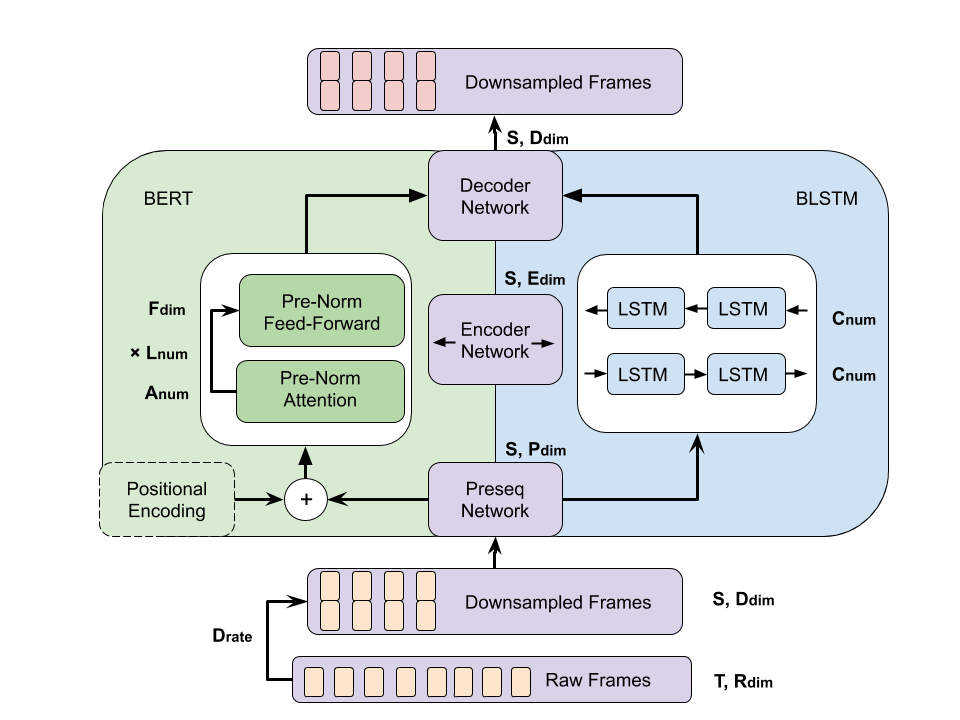}
\caption{Overview of proposed framework. On the left: BERT encoder in green shade. On the right: BLSTM encoder in blue shade.}
\label{fig:V1}
\end{figure}

Figure \ref{fig:V1} is the system diagram of the proposed approach. The system is composed of three modules: pre-sequence network, encoder network and decoder network, shown in purple shade. The encoders in comparison are plotted in parallel, shown in green and blue shades. A summary of hyper-parameters is shown in Table \ref{table: hp}. 

The system input is raw frames $R \in \mathbb{R}^{T \times R_{dim}}$, where $T$ is the total number of frames, and $R_{dim}$ is the dimension of log Mel features. To handle very long input raw frames, adjacent $D_{rate}$ frames of $R$ are stacked along frequency axis to form a downsampled sequence of length $S$ \cite{sperber2018self}. 

The pre-sequence network extracts local spectral and temporal features before the sequence is fed into the backbone sequence model. CNN2d captures local pictorial features, such as pitch contour. CNN1d performs frequency-wise filtering across time. (B)LSTM provides order to higher level layers \cite{sperber2018self}. DEF projects spectral features to hidden representations that is suitable for addition with positional encoding \cite{liu2020mockingjay}. All pre-sequence models produce the same hidden dimension $P_{dim}$.

Since the BERT architecture is invariant to order, CNN2d, CNN1d and DEF pre-sequence networks are all added with sinusoidal positional encoding before feeding into the BERT encoder. The encoder networks extract high level semantic information. Hyperparameters of BERT and BLSTM are chosen to keep model size similar. The transformer layer follows a prenorm \cite{wang2019learning} architecture to facilitate gradient flow. 

The decoder network maps the encoded hidden representations back to stacked log Mel features $\hat{D} \in \mathbb{R}^{S \times D_{dim}}$. $\ell_2$ loss is calculated in the end. 

\begin{table}[tb]
\centering
\caption{\textsl{Hyper-parameters. FC-n stands for fully connected layer of size n. LSTM shows per direction cells. CNN shows numKernels-kernelShape-stride-padding.}}
\vspace{0.5em} 
\label{table: hp}
\resizebox{\columnwidth}{!}{%
\begin{tabular}{@{}l|l|l|l@{}}
      \toprule
      \multicolumn{3}{l|}{$R_{dim}=80, D_{rate}=3, D_{dim}=240, P_{dim}=768$}                                                  & \# params \\ \midrule
      \multirow{5}{*}{Preseq} & DEF   & FC-768                               & 0.184M \\ \cmidrule(l){2-4} 
                               & CNN2d & \begin{tabular}[c]{@{}l@{}}CNN: 64-(11, 10)-(1, 5)-(5, 0)\\ FC-768\end{tabular}       & 2.317M    \\ \cmidrule(l){2-4} 
                              & CNN1d & CNN: 768-11-1-5                      & 2.027M \\ \cmidrule(l){2-4} 
                              & LSTM  & $C_{num}$: 384                       & 1.923M \\ \cmidrule(l){2-4} 
                              & CL    & CNN2d preseq $\rightarrow$ LSTM preseq & 3.527M \\ \midrule
      \multirow{2}{*}{Encoder} & BERT  & \begin{tabular}[c]{@{}l@{}}$A_{num}=16, F_{dim}=2048, L_{num}=3$\\ $E_{dim}$: 768\end{tabular} & 16.54M    \\ \cmidrule(l){2-4} 
                               & BLSTM & \begin{tabular}[c]{@{}l@{}}$C_{num}$: 1024\\ $E_{dim}$: 2048\end{tabular}             & 14.69M    \\ \midrule
      \multicolumn{2}{l|}{Decoder}    & FC-256 $\rightarrow$ Relu $\rightarrow$ FC-240          & 0.260M \\ \bottomrule
      \end{tabular}%
      }
      \end{table}

\section{Experiments}
\label{sec:exps}

\begin{table}[tb]
\centering
\caption{\textsl{PESQ scores, number of parameters and inference times. Inference time is reported with 330 test utterances with a batch size of 10. }}
\vspace{0.5em}
\label{table: pesq}  
\resizebox{\columnwidth}{!}{%
\begin{tabular}{@{}ccccccc@{}}
\toprule
      & $T_{60}$ & CNN2d & CL         & CNN1d         & LSTM   & DEF  \\
      \cmidrule(l){2-7} 
BERT  & 0.3      & \textbf{1.971} & 1.886      & 1.903         & 1.912 & 1.890 \\
      & 0.6      & 1.733 & \textbf{1.736}      & 1.729         & 1.721 & 1.643 \\
      & 0.9      & 1.574 & \textbf{1.580}      & 1.576         & 1.554 & 1.469 \\ 
      \cmidrule(l){2-7} 
      & M        & 19.12 &\textbf{20.3}        & 18.82         & 18.72 & 16.90\\
      & ms       & 310.2 & 416.2      & 374.7         &\textbf{417.9} &417.0\\
      \cmidrule(l){2-7} 
\multicolumn{1}{l}{} & \multicolumn{1}{l}{$T_{60}$} & \multicolumn{1}{l}{CNN2d} & \multicolumn{1}{l}{CL} & \multicolumn{1}{l}{CNN1d} & \multicolumn{1}{l}{WPE} & \multicolumn{1}{l}{NOISY} \\
BLSTM & 0.3      & \textbf{2.013} & 1.860 & \multicolumn{1}{c|}{1.991} & 1.541 & 1.448 \\
      & 0.6      & \textbf{1.793} & 1.731 & \multicolumn{1}{c|}{1.793} & 1.273 & 1.248 \\
      & 0.9      & \textbf{1.651} & 1.599 & \multicolumn{1}{c|}{1.634} & 1.210 & 1.191 \\
      \cmidrule(l){2-7} 
      & M         & 17.60& \textbf{18.2} &  \multicolumn{1}{c|}{17.31}     & --   & -- \\
      & ms         & 314.9& \textbf{495.6} &  \multicolumn{1}{c|}{309.2}     & --   & --
\end{tabular}%
}
\end{table}

\begin{table}[tb]
\centering
\caption{\textsl{PESQ of CNN1d-BERT with Griffin-Lim 32 iterations (GL) and Neural vocoder (NV)}}
\label{table: gl}
\resizebox{\columnwidth}{!}{%
\begin{tabular}{@{\hskip 0.1in}cc@{\hskip 0.1in}c@{\hskip 0.1in}c@{\hskip 0.1in}}
\toprule
$T_{60}$       & 0.3   & 0.6   & 0.9   \\ \midrule
GL/NV    & 1.941/1.903 & 1.752/1.729 & 1.592/1.576 \\
 \bottomrule
\end{tabular}%
}
\end{table}

\begin{figure*}[tb]
\centering
\subfigure[]{\includegraphics[width=0.33\textwidth]{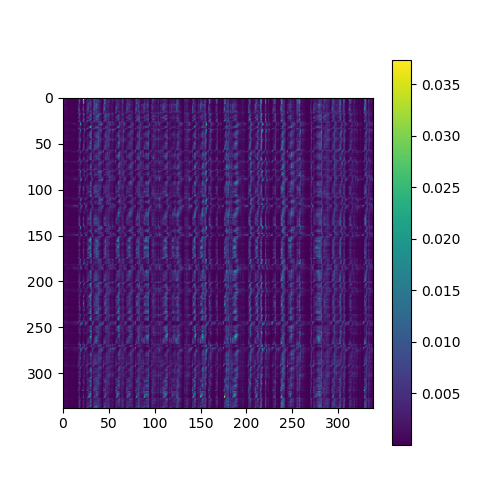}} 
\subfigure[]{\includegraphics[width=0.33\textwidth]{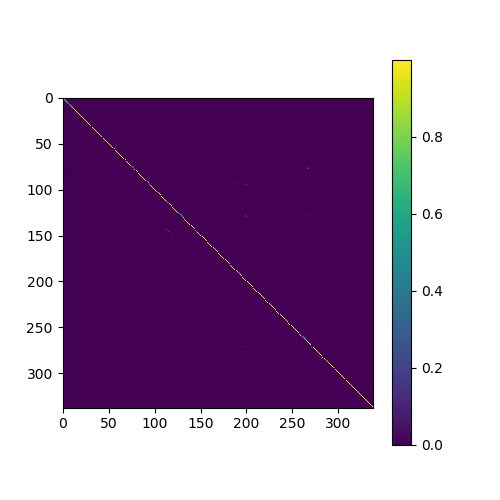}} 
\subfigure[]{\includegraphics[width=0.33\textwidth]{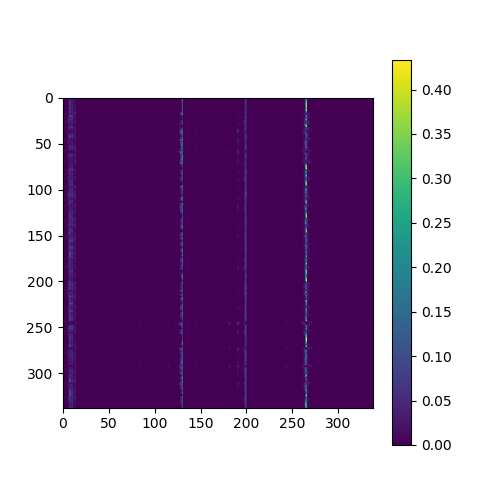}}
\caption{\textit{Attention score patterns. (a) global (b) diagonal (c) vertical}}
\label{fig:att}

\end{figure*}

\begin{figure}[tb]
\centering
\includegraphics[width=\columnwidth]{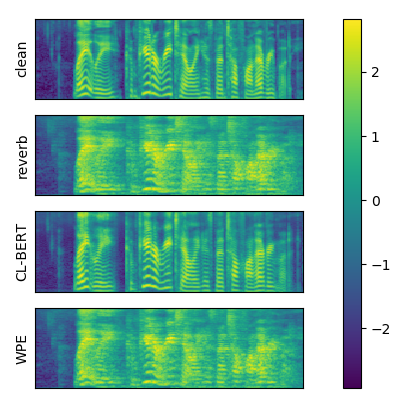}
\label{fig:spec}
\caption{\textsl{Mel-spectrograms before and after dereverberation. }}
\end{figure}

\subsection{Dataset}
To evaluate the performance of the proposed method, we create a synthetic dataset. The synthetic dataset is composed of 2 parts: clean speech files and synthetic room impulse responses (RIRs). The clean speech files are taken from the 3rd CHiME challenge corpus. 7138 utterances from 'tr05\_org' are used for training; 410 utterances from 'dt05\_bth' are used for validation; 330 utterances from 'et05\_bth' are used for testing. The RIRs are generated using the implementation of image method in \cite{lehmann2008prediction}. The simulation is configured as a room of size [4, 4, 2.5] meters, with sound source placed in the center of the room, and a single sound receiver placed 1 meters away at the same height with sound source. We use 3 reverberation times $T_{60}$: 0.3s, 0.6s, 0.9s. For each reverberation time, 11 RIRs are generated by sampling the circle of 1 meter radius uniformly around the sound source; 10 RIRs are used for training and validation, 1 unseen RIR is used for testing. Each clean speech in training/validation/testing set is convolved with a random RIRs uniformly from its candidate RIRs pool, i.e. 3 * 10 for training/validation and 3 * 1 for testing.

\subsection{Experiment Setups}
We align the preprocessing recipe with Parallel WaveGan for LibriTTS corpus to use their pre-trained vocoder \footnote{\url{https://github.com/kan-bayashi/ParallelWaveGAN}}. The sampling rate of the 3rd CHiME corpus is resampled to 24 kHz. We apply short-time Fourier transform (STFT) with a Hanning window of 1200 points, Fast Fourier transform (FFT) of 2048 points and a hop size of 300 points. Mel-scale spectrogram of 80 bins is then extracted over 80 to 7600 Hz. The Mel-features is normalized to zero mean and unit variance. Reconstructed waveforms are resampled to 16 kHz before perceptual evaluation of speech quality (PESQ) is performed. 

We compare our approach with weighted prediction error (WPE) \cite{drude2018nara} and BLSTM based models. WPE is a statistical method that removes only late reverberation $y_{late} = h_{late} \ast x$ from $y$, where $h_{late}$ is usually taken as the tail of $h$ after first $50ms$ \cite{naylor2010speech}. BLSTM based systems achieve state-of-the-art in speech dereverberation and denoising \cite{tan2018convolutional,zhao2018convolutional}. We train both BERT and BLSTM based systems from scratch. All networks are initialized with normal distribution and trained with Adam optimizer. The learning rate is warmed up over the first 1\% of the total 75k steps to a maximum of 3e-4 and then linearly decayed to 0. 

Three comparison experiments are summarized in Table \ref{table: pesq}. We first compare the performance of WPE with BLSTM based methods and a basic BERT (DEF-BERT) based method. It is clear that DEF-BERT outperforms WPE consistently and achieves comparable performance with CL-BLSTM method with less parameters (16.9M vs 18.2M).

The second experiment compares pre-sequence layers. We evaluated LSTM-BERT, CNN2d-BERT, CNN1d-BERT, CL-BERT with DEF-BERT. It is shown that convolutional and recurrent pre-sequence layers achieve consistently better performance than DEF-BERT in all except one case (DEF vs CL-BERT, 0.3s). Pre-sequence layer achieves more advantage as reverberation time increases, especially for convolutional pre-sequence layer, which is more capable of capturing local patterns over pure LSTM ones.

The third experiment compares BLSTM-based sequence models with BERT-based sequence models using the same pre-sequence layers CNN2d, CL and CNN1d. From Table \ref{table: pesq}, it is shown that BERT-based methods achieve comparable performance in all cases with slightly more parameters. BERT benefits from a stack of CNN2d and LSTM more than BLSTM, probably due to BERT favoring positional information. 

We then compare phase reconstruction methods in Table \ref{table: gl}. Griffin-Lim with 32 iterations obtains an real time factor (RTF) of 0.319, whereas Parallel WaveGan on GPU obtains an RTF of 0.027. Decoding with Parallel WaveGan is about 10 times faster than Griffin-Lim with comparable performance. However, decoding using a neural vocoder is rarely considered in the context of speech dereverberation and denoising. 

Figure ~\ref{fig:spec} shows an example utterance from the test set. The top row depicts spectrogram of the clean speech. The second row depicts reverberated speech at $T_{60}=0.6s$. The spectrogram is apparently expanded and blurred from convolution. On the bottom row, spectrogram after dereverberation with WPE is shown. WPE only removes late reverberation, therefore the enhanced spectrogram is still smeared by early reverberation. The third row demonstrates that CL-BERT can handle both early and late reverberation properly. The clean frame is disentangled from a superposition of copies of neighboring frames.  

Inspired by \cite{yang2020understanding}, we don't constrain frames to be only able to attend to neighboring frames. In Figure \ref{fig:att}, 3 patterns reported in \cite{yang2020understanding} are also found in the proposed method. This demonstrates that \{preseq\}-BERT can use frames farther apart to perform dereverberation. We validate this assumption with a consecutive masking training recipe \cite{liu2020mockingjay}. Experiment shows that \{preseq\}-BERT is able to recover raw frame at time $t$ with input frames from $t-4$ to $t+4$ masked out.

\section{Conclusions}
In this work, we have demonstrated the effectiveness of BERT as a backbone sequence model for speech dereverberation. Several pre-sequence layers have been proved to be beneficial for adapting BERT from NLP to speech. We also demonstrate that neural vocoder is effective in implicit phase reconstruction in the context of speech dereverberation. 

\clearpage
\bibliographystyle{IEEEbib}
\bibliography{library}

\end{document}